\documentclass[%
 reprint,
 amsmath,amssymb,
 aps,
]{revtex4-2}

\usepackage{graphicx}
\usepackage{dcolumn}
\usepackage{color}
\usepackage{bm}

\usepackage{color,amsmath}

\usepackage{graphicx}

\newcommand{\Br}{{\bf r}}

\newcommand{\Be}{{\bf e}}

\newcommand{\BE}{{\bf E}}

\newcommand{\Tr}{{\rm Tr}}

\newcommand{\xhat}{{\bf \hat{x}}}
\newcommand{\yhat}{{\bf \hat{y}}}

\newcommand{\ket}[1]{\left| #1 \right\rangle}
\newcommand{\bra}[1]{\left\langle #1 \right|}
\newcommand{\braket}[2]{\left\langle #1 \right|\left. #2 \right\rangle}

\begin{document}

\preprint{APS/123-QED}

\title{A unified matrix representation for spin and orbital angular momentum in partially coherent beams}

\author{Olga Korotkova}
\email{korotkova@physics.miami.edu}
 \affiliation{Department of Physics, University of Miami, Coral Gables, FL 33146, USA}
\author{Greg Gbur}%
 \email{gjgbur@uncc.edu}
\affiliation{Department of Physics and Optical Science, The University of North Carolina at Charlotte, Charlotte, North Carolina 28223, USA}%

\date{\today}

\begin{abstract}
A matrix representation is introduced for stationary beam-like light fields that allows for simultaneous treatment of the orbital angular momentum (OAM) and the second-order state of spatial coherence. This Coherence-OAM matrix is a counterpart of the  $2\times2$ beam Coherence-polarization matrix in a Cartesian polarization basis or Coherence-spin angular momentum (SAM) matrix in a circular polarization basis. The general properties of the Coherence-OAM matrix are discussed and then the Coherence-SAM-OAM matrix is introduced as a unifying entity for treating beams with any coherence, spin and orbital angular momentum states. 
\end{abstract}

\maketitle


\section{Introduction}

Over the past three decades, there has been an intense amount of interest in the properties and applications of light beams carrying orbital angular momentum  \cite{1,2}. These beams have been considered for optical communications \cite{3,4}, for the trapping and rotation of small particles \cite{5,6}, and for image processing \cite{7,8}, among other applications. The study of beams carrying orbital angular momentum (OAM) is a subset of the field of singular optics, which is concerned with the topological singularities in various properties of light fields \cite{9}. Laguerre-Gauss beams are commonly used pure states of OAM; they possess a line of zero intensity on the propagation axis with a helical phase around it, usually referred to as an optical vortex, and the OAM per photon is directly proportional to the axial phase twist.  For non-pure states, the relationship between the phase structure and the OAM is much more complicated \cite{10}.

The introduction of partial coherence to OAM beams at first may appear paradoxical, as OAM is usually encoded in the phase structure of light and a beam with partial spatial coherence by definition does not have a well-defined phase. However, there has been an effort over a number of years to understand the relationship between coherence and OAM; this includes the study of how optical vortices become singularities of the correlation function as spatial coherence is decreased \cite{11}-\cite{13} as well as investigations into the OAM characteristics of partially coherent beams in general \cite{14}.  Partially coherent vortex beams are now being considered for use in free-space optical communications \cite{15,16}, as it is known that random beams tend to be less distorted by atmospheric turbulence than their fully coherent counterparts \cite{17}.

It is now known that there are richer varieties of OAM states in partially coherent fields. It is possible to impart a pure phase twist on a partially coherent beam \cite{18}, and in twisted Gaussian Schell-model beams the twist is contained entirely in the statistical properties of the field \cite{19}. These two possibilities have recently been combined into twisted vortex Gaussian Schell-model beams \cite{20}, showing the potential of mixing coherence and OAM.

As yet, however, it does not appear that anyone has introduced a formalism to describe and study the interplay of coherence and OAM in partially coherent beams. Here we may draw an analogy to work that has been done on the unified theory of coherence and polarization \cite{21}, in which a cross-spectral density matrix (the Coherence-polarization matrix) is introduced to simultaneously analyze the effects of coherence and polarization on the observable properties of the field. If this matrix is written in a circular polarization basis instead of a linear polarization basis, it represents spin states of the beam and may be referred to as the Coherence-spin angular momentum (SAM) matrix. The Coherence-polarization matrix has become a standard tool for the study of vector partially coherent beams \cite{22,23}.

With this in mind, it is natural to wonder whether one can make an analogous matrix representation of a partially coherent beam that decomposes it into its orbital angular momentum states. In this paper we introduce such a Coherence-OAM matrix (COAM), describe its properties, and consider its similarities and differences with the Coherence-SAM matrix (CSAM). We describe a novel experimental arrangement for measuring elements of this matrix, and introduce a number of examples.  We end by generalizing this matrix to include polarization effects as well, to produce a general Coherence-SAM-OAM matrix.

We begin with a review of the relevant aspects of coherence theory in the space-frequency domain, then introduce the new matrix and its properties.

\section{Review of coherence theory}

In modern optical coherence theory, a random scalar light field is commonly characterized by the cross-spectral density (CSD), a second-order correlation function \cite{24},
\begin{equation}\label{CSD}
W(\Br_1,\Br_2;\omega)=\langle U^\ast(\Br_1;\omega) U(\Br_2;\omega) \rangle_\omega, 
\end{equation}
where the angular brackets represent an ensemble average over a set of monochromatic realizations of the scalar optical field $U(\Br;\omega)$, where $\Br$ is a position vector, $\omega$ is the angular frequency, and the star denotes complex conjugation. The CSD satisfies a pair of Helmholtz equations and it contains complete information about the spectral density and the degree of coherence of the field, given respectively by the expressions
\begin{equation}\label{S}
S(\Br;\omega)=W(\Br,\Br;\omega),
\end{equation}
\begin{equation}\label{mu}
\mu(\textbf{r}_1,\textbf{r}_2;\omega)=\frac{W(\textbf{r}_1,\textbf{r}_2;\omega)}{\sqrt{S(\textbf{r}_1;\omega)}\sqrt{S(\textbf{r}_2;\omega)}}.
\end{equation}
  
Much of the early work in coherence theory (though not all) focused on scalar waves and effectively neglected polarization. But the work of James \cite{25}, Gori \cite{26} and Wolf \cite{27} highlighted that the combination of coherence and polarization produces non-trivial effects, such as changes of the state and degree of polarization as a wave propagates. In this case, the CSD can be generalized to a $2\times2$ matrix, known as the CSD matrix.

To derive this matrix, we now consider an electric field vector $\BE(\Br;\omega)$ which is transverse to the overall beam propagation direction $z$.  We may decompose this field into a particular basis set of transverse unit vectors $\Be_\alpha$ through the use of the dot product,
\begin{equation}
E_\alpha(\Br;\omega) = \BE(\Br;\omega)\cdot \Be_\alpha.\label{E:projection}
\end{equation}
For example, we may use a Cartesian basis where $\alpha = \{x,y\}$ in which case $\Be_x = \xhat$ and $\Be_y = \yhat$, or we may choose a circular polarization basis where $\alpha = \{+,-\}$, in which case $\Be_\pm = (\xhat\pm i\yhat)/\sqrt{2}$. Here $\Be_+$ and $\Be_-$ represent left and right circular polarization, respectively.

The cross-spectral density matrix is then constructed from the ensemble average of pairs of components in the chosen basis,
\begin{equation}
W_{\alpha \beta}(\Br_1,\Br_2,\omega) =\langle E^\ast_\alpha(\Br_1;\omega)E_\beta(\Br_2;\omega)\rangle_\omega.\label{SAM:elements}
\end{equation}

For comparison with the OAM case, it is worthwhile to note that we may also consider the matrix as being derived from the cross-spectral density dyadic, constructed from the direct product of electric field vectors without specifying a basis,
\begin{equation}\label{W:outer}
\overleftrightarrow{W}(\Br_1,\Br_2,\omega) = \langle \BE^\dagger(\Br_1;\omega)\otimes \BE(\Br_2;\omega)\rangle_\omega.
\end{equation}
where $\otimes$ represents the direct product.  Then left and right multiplication by basis vectors represents this dyadic in the chosen basis,
\begin{equation}
W_{\alpha \beta}(\Br_1,\Br_2,\omega)= \Be_\alpha^\dagger \cdot \overleftrightarrow{W}(\Br_1,\Br_2,\omega) \cdot \Be_\beta.
\end{equation}

For electromagnetic (EM) beams, the spectral density and the degree of coherence are defined as
\begin{equation}\label{Strace}
S(\Br;\omega)=\Tr \overleftrightarrow{W}(\Br,\Br;\omega),
\end{equation}
\begin{equation}
\mu(\Br_1,\Br_2;\omega)=\frac{\Tr \overleftrightarrow{W}(\Br_1,\Br_2;\omega)}{\sqrt{S(\Br_1;\omega)}\sqrt{S(\Br_2;\omega)}},
\end{equation}
where $\Tr$ stands for the matrix trace, and these expressions reduce to definitions (\ref{S}) and (\ref{mu}) in the scalar approximation. Moreover, individual elements of the CSD matrix can be measured experimentally through the Stokes parameters, and polarization properties of the field can be determined directly from the matrix, such as the degree of polarization:
\begin{equation}
\begin{split}\label{Psam}
P(\Br;\omega)&=\frac{S_{pol}(\Br;\omega)}{S(\Br;\omega)} \\&
=\frac{\lambda_1(\Br;\omega)-\lambda_2(\Br;\omega)}{\lambda_1(\Br;\omega)+\lambda_2(\Br;\omega)},
\end{split}
\end{equation}
where $S_{pol}$ is the spectral density of the polarized portion of the beam and $\lambda_1$, $\lambda_2$ are the larger and the smaller eigenvalues of $\overleftrightarrow{W}$. These eigenvalues are be real and non-negative due to the Hermiticity and non-negative definiteness of the matrix at a single position $\Br$.

If we choose to represent the matrix in a circular polarization basis, the eigenvalues $\lambda_\alpha(\Br;\omega)$ represent the intensities of the left and right circularly polarized components of the field. In this case, we may consider it a Coherence-SAM matrix.  Then the field is in a pure state of SAM at a position $\Br$ only if one of the eigenvalues is zero, indicating that the field is circularly polarized at that position.

\section{Definition of the Coherence-OAM matrix}

Just as we are able to decompose a general partially coherent vector field into a matrix of SAM states, we now consider the decomposition of a general partially coherent scalar field into a matrix of OAM states.

\begin{figure}
\centering
\includegraphics[width=8.5cm]{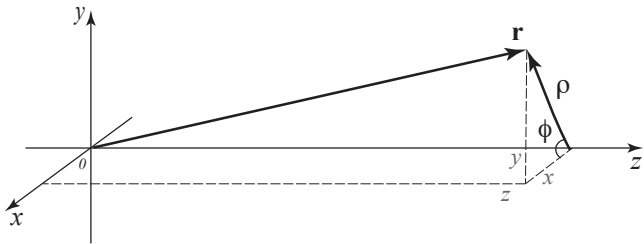}
\caption{Notations for beam propagation.} \label{notations}
\end{figure}

Let us consider a realization $U(\Br;\omega)$ of a beam-like scalar optical field, propagating from plane $z=0$ into the positive half-space $z>0$  (see Fig. \ref{notations}). We express the position vector $\Br$ in the cylindrical coordinate system $\Br=(\rho,\phi,z)$ and denote by $\pmb{\rho}=(\rho,\phi)$ the projection of $\Br$ onto a plane orthogonal to $z$ for brevity. For convenience we will use primes for vectors in the source plane $z=0$. 

To produce a Coherence-OAM matrix, we follow a procedure analogous to that used to derive the Coherence-SAM matrix. We imagine decomposing the field $U(\Br;\omega)$ into a basis set of azimuthal functions, written as $\ket{l}$ in bra-ket notation, where $l$ is an index labeling the mode. We then have
\begin{equation}\label{fielddecopm1}
\ket{U(\Br;\omega)} = \sum_{l=-\infty}^\infty U_l(\rho,z;\omega)\ket{l}.
\end{equation}

These states may be pure states of OAM, with $\braket{\phi}{l} = \exp[il\phi]$, or they could be trigonometric states, i.e. $\braket{\phi}{l_+}= \cos(l\phi)$ and $\braket{\phi}{l_-}=\sin(l\phi)$. The former case will produce the Coherence-OAM matrix, where the latter case may be said to produce a Coherence-trigonometric matrix. The elements $U_l(\rho,z;\omega)$ can be found by the inner product, i.e.
\begin{equation}
U_l(\rho,z;\omega) = \braket{l}{U(\Br;\omega)},
\end{equation}
which is the analogy to Eq.~(\ref{E:projection}), and we define a general Coherence-OAM matrix as the ensemble average of these elements at two points, i.e.
\begin{equation}\label{Wlm:def}
W_{lm}^{(OAM)}(\rho_1,\rho_2,z_1,z_2;\omega) \equiv \langle U_l^\ast(\rho_1,z_1;\omega)U_m(\rho_2,z_2;\omega)\rangle_\omega,
\end{equation}
which is the analogy to Eq.~(\ref{SAM:elements}).

The cross-spectral density in Eq.~(\ref{CSD}), is itself comparable to the cross-spectral density dyadic of Eq.~(\ref{W:outer}), as it represents the direct product in functional space. With this in mind, we may write the cross-spectral density in the operator form,
\begin{equation}
\begin{split}
\overleftrightarrow{W}(\Br_1,&\Br_2;\omega) \\& =\sum\limits_{l',m'=-\infty}^{\infty} W_{l'm'}^{(OAM)}(\rho_1,z_1,\rho_2,z_2;\omega)\ket{m'}\bra{l'},
\end{split}
\end{equation}
with
\begin{equation}
W_{l'm'}^{(OAM)}(\rho_1,\rho_2,z_1,z_2;\omega) = \bra{m'}\overleftrightarrow{W}(\Br_1,\Br_2;\omega)\ket{l'}.
\end{equation}

The preceding formalism is presented to show the analogy between the Coherence-polarization and Coherence-OAM matrices.  Let us now focus on the pure OAM states, for which the inner product is defined as
\begin{equation}\label{FTcoefficients}
\braket{l}{U(\Br;\omega)}=U_l(\rho,z;\omega)=\frac{1}{2\pi} \int_0^{2\pi} U(\rho,\phi,z;\omega)e^{-il\phi}d\phi.
\end{equation}
For these states, the functions $U_l(\rho,z;\omega)$ are simply the azimuthal Fourier series coefficients.

We may use this decomposition twice in Eq.~(\ref{Wlm:def}), to get an expression of the form,
\begin{equation}
\begin{split}
W(\Br_1,&\Br_2;\omega) \\& = \sum\limits_{l,m=-\infty}^{\infty}\langle U_l^\ast(\rho_1,z_1;\omega)U_m(\rho_2,z_2;\omega)\rangle_\omega e^{-i(m-l)\phi}.
\end{split}
\end{equation}

The individual elements of the COAM matrix can be found directly from the cross-spectral density by using \eqref{FTcoefficients} in \eqref{Wlm:def}, interchanging the order of integration and statistical averaging and applying \eqref{CSD}:
\begin{equation}\label{CSDdecomp4}
\begin{split}
W_{lm}^{(OAM)}&(\rho_1,z_1,\rho_2,z_2;\omega)= \langle U_l^*(\rho_1,z_1;\omega) U_m(\rho_2,z_2;\omega)\rangle_{\omega} \\&
=\frac{1}{(2\pi)^2}  \biggl< \int\limits_0^{2\pi}  [U(\textbf{r}_1;\omega)e^{-il \phi_1}]^* d\phi_1  \\& \times \int\limits_0^{2\pi}  U(\textbf{r}_2;\omega)e^{-im \phi_2} d\phi_2  \biggr>_{\omega} \\&
=\frac{1}{(2\pi)^2} \int\limits_0^{2\pi}\int\limits_0^{2\pi} e^{il\phi_1} W(\textbf{r}_1,\textbf{r}_2;\omega)  e^{-im\phi_2}d\phi_1 d\phi_2, \\&
\quad   \quad\quad\quad\quad\quad\quad\quad\quad\quad   -\infty <l,m< \infty.
\end{split}
\end{equation}

The original cross-spectral density may be recovered from the matrix elements through the expression
\begin{equation}\label{CSDdecomp}
\begin{split}
W(\Br_1,\Br_2;\omega)& = \sum_{l=-\infty}^\infty\sum_{m=-\infty}^\infty W_{lm}^{(OAM)}(\rho_1,z_1,\rho_2,z_2;\omega)\\& \times e^{-il\phi_1}e^{im\phi_2}.
\end{split}
\end{equation}

One important difference between the COAM matrix and the CSAM matrix is immediately evident.  The SAM is determined by the state of polarization, which is an independent degree of freedom from the spatial characteristics of the beam.  OAM, however, is determined by the spatial characteristics of the beam and is therefore dependent upon it. This causes significant differences in the way that important characteristics of the OAM matrix are analyzed and measured, as we will see.

\section{Basic properties of the COAM matrix}

\label{basicproperties}

A number of properties of the Coherence-OAM matrix may be derived by applying the known properties of the cross-spectral density. Here we consider basic properties of the matrix.

1) \textit{Dimensionality}. As follows from the Fourier series decomposition of Eq.~(\ref{CSDdecomp4}), the COAM matrix is a square, infinite-dimensional matrix. However, experimental measurements of OAM are limited to a finite range (on the order of $-10\leq l\leq 10$ in many experiments \cite{3,28}), so the matrix of accessible elements will be finite in practice. If the highest and the lowest indices at which the Fourier components are non-trivial are $l_{min}$ and $l_{max}$, respectively, then the COAM matrix has dimension $(l_{max}-l_{min}+1)\times (l_{max}-l_{min}+1)$.   

2) \textit{Quasi-Hermiticity}. The cross-spectral density $W(\Br_1,\Br_2;\omega)$ is Hermitian, which in turn places conditions on the behavior of elements of the Coherence-OAM matrix. Starting with the definition of Hermiticity of the cross-spectral density, 
\begin{equation}\label{QH}
 W^*(\textbf{r}_1,\textbf{r}_2;\omega)=W(\textbf{r}_2,\textbf{r}_1;\omega),
\end{equation}
we find the following relationship for the matrix elements,
\begin{equation}\label{W:Herm}
\begin{split}
W_{lm}^{(OAM)*}&(\rho_1,z_1,\rho_2,z_2;\omega) \\&  =\frac{1}{(2\pi)^2} \int\limits_0^{2\pi}\int\limits_0^{2\pi} e^{-il\phi_1} W^*(\textbf{r}_1,\textbf{r}_2;\omega)  e^{im\phi_2}d\phi_1 d\phi_2 \\& 
 =\frac{1}{(2\pi)^2} \int\limits_0^{2\pi}\int\limits_0^{2\pi} e^{-il\phi_1} W(\textbf{r}_2,\textbf{r}_1;\omega)  e^{im\phi_2}d\phi_1 d\phi_2 \\&
=W_{ml}^{(OAM)}(\rho_2,z_2,\rho_1,z_1;\omega).
\end{split}
\end{equation}
This implies a quasi-Hermiticity for the matrix elements, which satisfy Hermiticity when both the position variables as well as the matrix indices are swapped. Though we have derived this expression for the OAM basis specifically, Eq.~(\ref{W:Herm}) can be shown to apply for any choice of azimuthal basis.

3) \textit{Non-negative definiteness} of the COAM matrix directly follows from that of the CSD function, i.e.,
\begin{equation}
\iint f^*(\textbf{r}_1;\omega)f(\textbf{r}_2; \omega) W(\textbf{r}_1,\textbf{r}_2;\omega) d\textbf{r}_1  d\textbf{r}_2 \geq 0,
\end{equation}
with $f(\Br;\omega)$ being an arbitrary complex-valued function and the integration extends twice over a selected domain. For beam-like fields, it is sufficient to establish such a condition across any transverse plane such as the source plane. Using the matrix representation  (\ref{CSDdecomp4}) of the CSD and decomposing the function $f$ into the polar Fourier basis,
\begin{equation}
f(\rho,\phi, 0;\omega)=\sum\limits_{l=-\infty}^{\infty} f_l(\rho,0;\omega)e^{il\phi},
\end{equation}
we find that the COAM matrix components must obey the inequality
\begin{equation}\label{W:NND}
\begin{split}
\sum\limits_{l,m=-\infty}^{\infty} \iint\limits_{0}^{\infty}& \rho_1\rho_2f_l^*(\rho_1,0;\omega)f_m(\rho_2,0; \omega) \\& \times W_{lm}^{(OAM)}(\rho_1,0,\rho_2,0;\omega) d\rho_1 d\rho_2 \geq 0.
\end{split}
\end{equation}
It is to be noted that the product $\rho_1\rho_2$ may be absorbed into the product $f_l^\ast f_m$ to simply the expression.  As in the case of quasi-Hermiticity, though we have done the calculation specifically for the OAM basis, Eq.~(\ref{W:NND}) can be shown to apply for any azimuthal basis.

4) A \textit{pure OAM state} is one which has a simple $\exp[im\phi]$ dependence. It is characterized by a COAM matrix with a single on-diagonal element,
\begin{equation}\label{PureStates}
W^{(OAM)}_{lm}(\rho_1,z_1,\rho_2,z_2;\omega)=\delta_{lp}\delta_{mp} W_0(\rho_1,z_1,\rho_2,z_2;\omega),
\end{equation} 
where $\delta$ stands for a Kronecker symbol and the function $W_0$ satisfies the Hermiticity and non-negative definiteness conditions given previously.  This is analogous to a circular polarization state in the Coherence-SAM matrix, in which case the $2\times 2$ matrix has a single non-zero diagonal element.

5) \textit{Coherent OAM state.} A field which is spatially coherent will have a factorized cross-spectral density, i.e.
\begin{equation}
W(\Br_1,\Br_2;\omega) = U^\ast(\Br_1;\omega)U(\Br_2;\omega).
\end{equation}
This implies in turn that the COAM matrix may be factorized as
\begin{equation}
\begin{split}\label{CoherentSates}
W^{(OAM)}_{lm}(\rho_1,z_1,&\rho_2,z_2;\omega)\\& =  U_{l}^{(OAM)*}(\rho_1,z_1;\omega) U_{m}^{(OAM)}(\rho_2,z_2;\omega).
\end{split}
\end{equation}
It is to be noted that being a pure OAM state indicates complete coherence in the azimuthal direction, but the field may be partially coherent in the radial direction. Being a coherent OAM state indicates that the field is spatially coherent, but does not imply that it is in a pure OAM state: the field may have any azimuthal behavior.

6) \textit{Correlated OAM state.} 
In another special case the individual COAM matrix elements may factorize as
\begin{equation}
W_{lm}^{(OAM)}(\rho_1,\rho_2)=I_{lm}U^*_{l}(\rho_1) U_{m}(\rho_2),  
\end{equation}
where $I_{lm}$ are coefficients that satisfy Hermiticity of the COAM matrix, i.e., $I_{lm}=I^*_{ml}$, $I_{ll}\geq 0$. Such a state can be regarded as a counterpart of a completely polarized light in the CSD matrix formulation. 

7) \textit{Parseval's identity}. Provided that $U(\textbf{r};\omega) \in L^2([0,2\pi])$, i.e., being square integrable with respect to $\phi$,   
\begin{equation}
\sum\limits_{l=-\infty}^{\infty}  W^{(OAM)}_{ll}(\rho,z,\rho,z;\omega)  =\frac{1}{2\pi}\int\limits_{0}^{2\pi} \bigl< |U(\rho,\phi,z)|^2 \bigr>_{\omega} d\phi. 
\end{equation} 
This formula implies that at any fixed plane $z$ and at any fixed radial position $\rho$ the energy carried by the beam is equal to the sum of all contributions contained in the pure OAM states (see property 4).

\section{Measurement of the COAM matrix elements} 
 
\begin{figure}
\centering
\includegraphics[width=8.5cm]{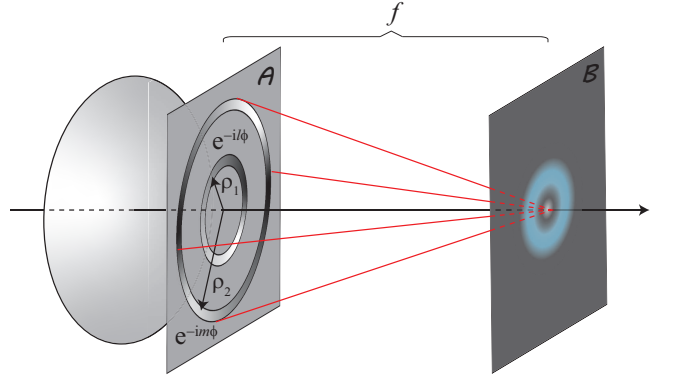}
\caption{Schematic diagram for measurement of the COAM matrix elements.}\label{YIEfigure}
\end{figure}

For the Coherence-OAM matrix to be of practical use, it must be a measurable quantity. In order to isolate the individual Coherence-OAM matrix elements, we propose a ``double-ring'' version of the Young's double-slit experiment (see Fig. \ref{YIEfigure}). The field to be measured impinges onto plane $\mathcal{A}$ at $z=0$ in which two narrow ring apertures, with coinciding centers, of radii $\rho_1$ and $\rho_2$ are present.  In the ring apertures, spiral plates with phase patterns $e^{-il\phi}$ and $e^{-im\phi}$ are inserted. A Gaussian lens with a focal length $f$ is then placed right before plane $\mathcal{A}$ with the center coinciding with that of the beam. The spectral density of interfering fields is then measured in the focal plane $\mathcal{B}$ at  $z=f$. 

The crux of the method: any component of the field that is transmitted through a ring aperture with an azimuthal phase will produce a donut-shaped intensity pattern in the focal plane, with a zero of intensity at the geometric focus. Due to the spiral plates, only those modes with azimuthal orders $l,m$ will produce a non-zero intensity at geometric focus: the rings isolate the radial position, while the spiral plates isolate azimuthal order.

If the field impinging onto plane $\mathcal{A}$ is $U(\pmb{\rho},0;\omega)$, then the transmitted field is approximately
\begin{equation}
\begin{split}
U_t(\pmb{\rho},0;\omega)&=U_t(\pmb{\rho},0;\omega)e^{-il\phi}\delta(\rho-\rho_1)\Delta \\& +U_t(\pmb{\rho},0;\omega)e^{-im\phi}\delta(\rho-\rho_2)\Delta,
\end{split}
\end{equation} 
where we have assumed the radial width $\Delta$ of the rings is sufficiently narrow to approximate the apertures using delta functions. The field diffracted to plane $\mathcal{B}$ becomes
\begin{equation}
U(\pmb{\rho},f;\omega)=\frac{e^{ikf}}{\lambda f}e^{ik\rho^2/2f} \int U_t(\pmb{\rho}',0;\omega) e^{-ik\pmb{\rho} \cdot \pmb{\rho}'/f} d\pmb{\rho}',
\end{equation}
where $k$ is the wave number, and, in particular, at the focal point $\rho=0$,
\begin{equation}\label{focal}
U(0,0,f;\omega)=\frac{e^{ikf}}{\lambda f} \int U_t(\pmb{\rho}',0;\omega)  d \pmb{\rho}'.
\end{equation}
Then taking the average of the field at the focal point in plane $\mathcal{B}$ and using \eqref{focal} yields
\begin{equation}
\begin{split}
&W(0,0,f;\omega) =\bigl< U^*(0,f;\omega)U(0,f;\omega) \bigr>_{\omega} \\&
=\frac{1}{\lambda^2f^2}\iint \bigl< U_t^*(\pmb{\rho}'_1,0;\omega) U_t(\pmb{\rho}'_2,0;\omega) \bigr>_{\omega} d\pmb{\rho}'_1 d\pmb{\rho}'_2\\&
=\frac{\Delta^2}{\lambda^2f^2}\Biggl\{ \rho_1^2 \iint e^{il\phi'_1} W(\rho_1,\phi'_1,0,\rho_1,\phi'_2,0;\omega)\\& \times e^{-il\phi'_2}d\phi'_1 d\phi'_2  \\&
+\rho_2^2 \iint e^{im\phi'_1} W(\rho_2,\phi'_1,0,\rho_2,\phi'_2,0;\omega)e^{-im\phi'_2}d\phi'_1 d\phi'_2 \\&+
\rho_1\rho_2 \iint e^{il\phi'_1} W(\rho_1,\phi'_1,0,\rho_2,\phi'_2,0;\omega)e^{-im\phi'_2}d\phi'_1 d\phi'_2 \\&
+  \rho_1\rho_2 \iint e^{im\phi'_1} W(\rho_2,\phi'_1,0,\rho_1,\phi'_2,0;\omega)e^{-il\phi'_2}d\phi'_1 d\phi'_2
\Biggr\},
\end{split}
\end{equation}
Applying the definition of the  COAM matrix elements given in \eqref{CSDdecomp4} and their quasi-Hermiticity expressed in \eqref{QH} results in the interference law
\begin{equation}\label{YIE}
\begin{split}
W^{(OAM)}&(0,0,f;\omega)=\frac{4\pi^2\Delta^2}{\lambda^2f^2}\Bigl\{  \rho_1^2W^{(OAM)}_{ll}(\rho_1,0,\rho_1,0;\omega)\\&+\rho_2^2W^{(OAM)}_{mm}(\rho_2,0,\rho_2,0;\omega)  \\&+2\rho_1\rho_2Re\left[ W^{(OAM)}_{lm}(\rho_1,0,\rho_2,0;\omega)\right]   \Bigr\}.
\end{split}
\end{equation}
This expression is analogous to the traditional spectral interference law for Young's double-slit experiment \cite{29}. Thus the spectral density of the beam pre-filtered with a pair of spiral plates and the two concentric rings, measured in the focal plane of the lens, provides access to the interference pattern of the individual OAM modes. First, the first two terms in \eqref{YIE} are obtained by closing each of the rings in turn and then the third term is evaluated as the difference between the total spectral density and those which originate from individual rings.  

It is to be noted that a practical experiment will measure the intensity over a finite region around the focal point, resulting in crosstalk between the measured components of the matrix. This can be reduced by using a mode multiplication scheme, in which azimuthal modes of the beam are mapped to twice their original order, or more \cite{30}. Since higher-order OAM modes have a wider intensity null in their center, this mapping will reduce the crosstalk.

The Coherence-SAM-OAM introduced later in Section \ref{EM:gen} can also be measured via such an arrangement, by introducing different polarization sensitive optical elements in each annular ring, in addition to the spiral plates.

\section{Wolf equations for the COAM matrix elements}

In free space, the elements of the COAM matrix can be propagated independently, and with less computational effort than propagating the full cross-spectral density at once. As shown in Appendix A, the components of the matrix satisfy a pair of equations,
\begin{equation}\label{WE}
\begin{split}
\Bigl[\frac{\partial^2}{\partial \rho_i^2}+\frac{1}{\rho_i}\frac{\partial}{\partial \rho_i}  & -\frac{p_i^2}{\rho_i^2}+2ik\frac{\partial}{\partial z_i}\Bigr] \\& \times w^{(OAM)}_{lm}(\rho_1,z_1,\rho_2,z_2;\omega)=0,
\end{split}
\end{equation}
where $i = 1,2$, $(p_1,p_2)= (l,m)$, and $w^{(OAM)}_{lm}$ represents the paraxial form of the COAM elements, as defined in Eq.~(\ref{W:parax}).

Equations (\ref{WE}) represent the Wolf equations (originally introduced in \cite{31}) for the COAM matrix elements in the paraxial approximation, describing the evolution of the correlation functions. These equations are solved for $w^{(OAM)}_{lm}$ using an integral transform method in Appendix A, giving the result
\begin{equation}\label{Huygens}
\begin{split}
&w^{(OAM)}_{lm}(\rho_1,z_1,\rho_2,z_2;\omega)
\\& =\int\limits_{0}^{\infty}\int\limits_{0}^{\infty} \Biggl[ \rho'_1 \rho'_2 w^{(OAM)}_{lm}(\rho_1,0,\rho_2,0;\omega)
\\& \times \int\limits_0^{\infty}\kappa_1 J_l(\kappa_1 \rho_1) J_l(\kappa_1 \rho'_1) \exp\left(i\frac{\kappa_1^2 z_1}{2k} \right) d\kappa_1 \\& \times
\int\limits_0^{\infty}\kappa_2 J_m(\kappa_2 \rho_2) J_m(\kappa_2 \rho'_2) \exp\left(-i\frac{\kappa_2^2 z_2}{2k} \right) d\kappa_2\Biggr] d\rho'_1d\rho'_2. 
\end{split}
\end{equation}

This expression can be simplified further by applying the second Weber's integral \cite{32},
\begin{equation}
\int\limits_0^{\infty} \xi e^{-\gamma \xi^2}J_{\nu}(\alpha \xi)J_{\nu}(\beta \xi)d\xi=\frac{1}{2\gamma}e^{ -\frac{\alpha^2+\beta^2}{4\gamma} } I_{\nu}\left(\frac{\alpha\beta}{2\gamma} \right),
\end{equation}
which is valid for $\nu>-1$, $Re[\gamma]>0$.  It is to be noted that, as applied in Eq.~(\ref{Huygens}), $Re[\gamma]=0$. However, as done in \cite{33}, one can treat the case $Re[\gamma]=0$ as a limiting case when the wave is propagating in a medium with absorption, for which $k = k_R+ik_I$, and consider the limit in which $k_I$ is arbitrarily small.  Then Eq.~(\ref{Huygens}) may be simplified to the form,
\begin{equation}\label{Huygens2}
\begin{split}
w&^{(OAM)}_{lm}(\rho_1,z_1,\rho_2,z_2;\omega)
\\& =\frac{k^2}{z_1z_2}\int\limits_{0}^{\infty}\int\limits_{0}^{\infty} \Biggl[ \rho'_1 \rho'_2 w^{(OAM)}_{lm}(\rho_1,0,\rho_2,0;\omega) \\& \times
\exp\left(\frac{-ik(\rho_1^2+\rho_1'^2)}{2z_1}\right)\exp\left(\frac{ik(\rho_2^2+\rho_2'^2)}{2z_2}\right) \\ & \times J_l\left(\frac{k\rho_1\rho_1'}{z_1}\right)J_m\left(\frac{k\rho_2\rho_2'}{z_2}\right) d\rho'_1d\rho'_2. 
\end{split}
\end{equation}
Eq. (\ref{Huygens2}) involves only two integrations for each matrix element, as opposed to four simultaneous integrations needed to propagate the cross-spectral density as a whole.

\section{Local properties of the COAM matrix}  

The spectral density $S$ given in \eqref{S}, expressed via the elements of the COAM matrix, can be determined by substituting from \eqref{fielddecopm1} and evaluating the result at the coinciding spatial arguments:
\begin{equation}\label{spect:def}
\begin{split}
S(\textbf{r};\omega)&= \Bigl< \sum\limits_{l=-\infty}^{\infty} U_l^*(\rho,z;\omega) e^{-i l\phi}  \\& \times \sum\limits_{m=-\infty}^{\infty} U_m(\rho,z;\omega) e^{+i m\phi}  \Bigr>_{\omega} \\&
= \sum\limits_{l,m=-\infty}^{\infty}  W_{lm}^{(OAM)} (\rho,z,\rho,z;\omega) e^{-i(l-m)\phi},
\end{split}
\end{equation}
where in the last step we have used the definition in \eqref{CSDdecomp4}. The double sum consists of two types of contributions: the diagonal elements $W_{ll}^{(OAM)}$ and off-diagonal elements $ W_{lm}^{(OAM)}$ $(l \neq m)$  contributions from which can be combined with the help of \eqref{QH} as
\begin{equation}
\begin{split}
S(\textbf{r};\omega)&= Tr[\overleftrightarrow{W}^{(OAM)} (\rho,z,\rho,z;\omega)] \\&+ \sum\limits_{l,m=-\infty}^{\infty} Re[ W_{lm}^{(OAM)} (\rho,z,\rho,z;\omega) e^{-i(l-m)\phi}]. 
\end{split}
\end{equation}
where $Re$ stands for the real part and $m<l$ for the second term. The presence of the off-diagonal elements is a striking difference from spectral density of the CSD matrix [see \eqref{Strace}]; in the polarization case the off-diagonal elements do not contribute.  This difference may again be traced to the fact that the OAM states are determined by the spatial characteristics of the beam, and not separable from them.

The OAM flux density at a point is defined as \cite{14},
\begin{equation}
M(\textbf{r};\omega) = \frac{\varepsilon_0}{k}Im[\partial_{\phi_2}W(\rho_1,\phi_1,z,\rho_2,\phi_2,z;\omega) ]_{r_1=r_2},
\end{equation}
where $\varepsilon_0$ is the electric permittivity of free space. With the COAM matrix, this becomes
 \begin{equation}
 \begin{split}
M&(\textbf{r};\omega) \\& = \frac{\varepsilon_0}{k}Re\left[ \sum\limits_{l,m=-\infty}^{\infty} m W_{lm}^{(OAM)} (\rho,z,\rho,z;\omega) e^{-i(l-m)\phi}   \right].
\end{split}
\end{equation} 
Separating diagonal and off-diagonal elements yields 
\begin{equation}\label{M:Wll}
\begin{split}
M(\textbf{r};\omega)& = \frac{\varepsilon_0}{k} \Biggl[ \sum\limits_{l=-\infty}^{\infty} lW_{ll}^{(OAM)}(\rho,z,\rho,z;\omega)\\& 
+  \sum\limits_{l,m=-\infty}^{\infty}Re [m W_{lm}^{(OAM)} (\rho,z;\omega) e^{-i(l-m)\phi}]
\Biggr], 
\end{split}
\end{equation}
where as before $m<l$ for the second term. Thus the OAM flux density is the linear combination of all the terms contributing to the spectral density with coefficients depending on $l$ and $m$.  

We may write Eq.~(\ref{YIE}) in the modified form,
\begin{equation}\label{YIE2}
\begin{split}
W&(0,0,f;\omega)=\frac{4\pi^2\Delta^2}{\lambda^2f^2}\Bigl\{  \rho_1^2W_{ll}^{(OAM)}(\rho_1,0,\rho_1,0;\omega)\\&+\rho_2^2W_{mm}^{(OAM)}(\rho_2,0,\rho_2,0;\omega) \\& +2\rho_1\rho_2\sqrt{W_{ll}^{(OAM)}(\rho_1,0,\rho_1,0;\omega)} \\& \times \sqrt{W_{mm}^{(OAM)}(\rho_2,0,\rho_2,0;\omega)} Re\left[ \mu_{lm}(\rho_1,0,\rho_2,0;\omega)\right]   \Bigr\},
\end{split}
\end{equation}
and define a degree of correlation $\mu_{lm}$ between OAM modes as
\begin{eqnarray}\label{OAMcorr}
&&\mu_{lm}(\rho_1,0,\rho_2,0;\omega) \equiv \nonumber\\ && \frac{ W_{lm}^{(OAM)}(\rho_1,0,\rho_2,0;\omega)}{\sqrt{W_{ll}^{(OAM)}(\rho_1,0,\rho_1,0;\omega)W_{mm}^{(OAM)}(\rho_2,0,\rho_2,0;\omega)}}.
\end{eqnarray}
This degree of correlation represents the maximum visibility of interference fringes possible when interfering modes at different radial points $\rho_1$ and $\rho_2$ with different orders $l$ and $m$.  This quantity, like the traditional spectral degree of coherence, is bounded such that $0\leq |\mu_{lm}|\leq 1$.

The degree of coherence of a scalar field is defined by expression \cite{21}
\begin{equation}\label{DOC}
\mu(\pmb{\rho}_1,z,\pmb{\rho}_2,z;\omega)=\frac{W(\pmb{\rho}_1,z,\pmb{\rho}_2,z;\omega)}{\sqrt{S(\pmb{\rho}_1,z;\omega)}\sqrt{S(\pmb{\rho}_2,z;\omega)}}.
\end{equation}
It can be expressed via the elements of the COAM matrix by substituting from Eqs. (\ref{CSDdecomp}) and (\ref{spect:def}) into \eqref{DOC}:
\begin{equation}\label{DOC2}
\begin{split}
\mu(\pmb{\rho}_1,z,&\pmb{\rho}_2,z;\omega)\\& =\Biggl[\sum\limits_{l,m=-\infty}^{\infty}  W_{lm}^{(OAM)} (\rho_1,z,\rho_2,z;\omega) e^{-i(l-m)\phi} \Biggr] \\& \times
\Biggl[\sum\limits_{l,m=-\infty}^{\infty}  W_{lm}^{(OAM)} (\rho_1,z,\rho_1,z;\omega) e^{-i(l-m)\phi} \Biggr]^{-1/2} \\& \times
\Biggl[\sum\limits_{l,m=-\infty}^{\infty}  W_{lm}^{(OAM)} (\rho_2,z,\rho_2,z;\omega) e^{-i(l-m)\phi}\Biggr]^{-1/2}.
\end{split}
\end{equation}
Thus, the degree of coherence consists of all self- and joint correlations of pairs of OAM modes all being normalized by the same factor, in contrast with the OAM correlations in \eqref{OAMcorr}.

\section{Integrated properties of the COAM matrix}  

Standard measurements of the OAM properties of a field consider its integrated characteristics, not point-by-point measurements. When the COAM matrix is integrated over a beam cross-section, its properties become analogous to those of the Coherence-polarization matrix.

The total integrated spectral density of the field may be written as
\begin{equation}\label{integrated}
I(z;\omega) = \int S(\Br',z;\omega)d^2r' = \sum_{l=-\infty}^\infty S_l(z;\omega),
\end{equation}
where $S_l(z;\omega)$ can be found by integrating Eq.~(\ref{spect:def}) to be
\begin{equation}\label{OAMspectrum}
S_l(z;\omega) = 2\pi \int_0^\infty  W_{ll}^{(OAM)} (\rho',z,\rho',z;\omega)\rho' d\rho'.
\end{equation}
In comparison with Eq.~(\ref{Strace}), one finds that the integrated spectral density acts similarly to the spectral density for the Coherence-polarization matrix, involving a sum over the diagonal elements of the matrix.  The total intensity of each OAM component of the field can therefore be determined directly from the diagonal elements.

Similarly, by integrating Eq.~(\ref{M:Wll}), we may find the integrated OAM flux density,
\begin{equation}
M(z;\omega) = \frac{\varepsilon_0}{k} \sum_{l=-\infty}^\infty lS_l(z;\omega),
\end{equation}
which also depends only on the diagonal elements of the COAM matrix.

With these definitions, we can also introduce a \textit{degree of OAM purity} analogous to the degree of polarization given by Eq.~(\ref{Psam}). A simple definition can be given as
\begin{equation}\label{purity}
P^{(OAM)}(z;\omega) = \xi_{max}(z;\omega)-\xi_{min}(z;\omega)
\end{equation}
where 
\begin{equation}
\xi_l(z;\omega) = \frac{S_{l}(z;\omega)}{\sum\limits_{m=-\infty}^{\infty} S_m(z;\omega)}
\end{equation}
are the integrated OAM spectrum components defined in \eqref{OAMspectrum}, normalized by the total integrated spectral density given in \eqref{integrated}. The quantities $\xi_{max}$, $\xi_{min}$ are the maximum and the minimum values in the set. In cases when infinitely many spectral components are present, the limiting value $\xi_{min}=0$ can be set, as provided the beam power is finite the series must converge.

This definition of purity is common in characterization of matrix operators in functional analysis, as the ratio of the operator's range to its strength, and was in fact also found to be useful for characterizing the three-dimensional degree of polarization \cite{34}. Notwithstanding a close similarity of definitions \eqref{Psam} and \eqref{purity}, the fundamental difference between them stems from the fact that $P$ is the local beam property while $P^{(OAM)}$ is the integrated property of the whole wavefront.

Because the COAM matrix generally has a larger number of components than the CSAM matrix, other definitions of purity may prove useful. One alternative may be introduced by invoking the concept of quantum mechanical \textit{von Neumann entropy}, defined as
\cite{35}
\begin{equation}
N=-Tr(\overleftrightarrow{D} \ln \overleftrightarrow{D} ),    
\end{equation}
where $\overleftrightarrow{D}$ is a density matrix, $Tr$ denotes the trace and $\ln$ denotes the (natural) matrix logarithm. If $\overleftrightarrow{D}$ is diagonalized and expressed in terms of its eigenvectors: $|l\rangle$, and eigenvalues $\beta_l$, as
\begin{equation}
\overleftrightarrow{D}  =\sum\limits_l \beta_l |l\rangle \langle l|,
\end{equation}
then the von Neumann entropy is
\begin{equation}
N=-\sum _{l}\beta _{l} \ln \beta _{l}.
\end{equation}

Since the integrated COAM matrix enjoys a decomposition analogous to the density matrix, the definition above can be directly applied to it. Thus the (spectral) OAM entropy becomes
\begin{equation}
N^{(OAM)}(z;\omega)=-\sum\limits_{l=-\infty}^{\infty}\xi_{l}(z;\omega) \ln \xi_{l}(z;\omega).
\end{equation}

In order to obtain a degree of purity $P^{(OAM)}$ conveniently constrained to  interval $[0,1]$, we apply the frequently used map
\begin{eqnarray}
P^{(OAM)}(z;\omega)&=&\exp[-N^{(OAM)}(z;\omega)]\nonumber \\ &=&\exp\left[\sum\limits_{l=-\infty}^{\infty}\xi_{l}(\omega) \ln \xi_{l}(z;\omega)\right] \nonumber \\ &=&\prod\limits_{l=-\infty}^{\infty} \xi_{l}(z;\omega)^{\xi_{l}(z;\omega) }.
\end{eqnarray}

For a pure state $P^{(OAM)}(z;\omega)=1$ and for the ``white noise'' state $P^{(OAM)}(z;\omega)=0$. For a mixed state of any $L$ pure states each having $\xi^{(OAM)}_{l}(z;\omega) =1/L$, with the rest of eigenvalues trivial, we have $P^{(OAM)}(z;\omega)=1/L$. 

\section{Examples}

In this section, we demonstrate the COAM matrix representation for a number of model sources, and suggest ways of generating novel sources through the use of the COAM matrix.

(A) \textit{Sources with separable phase}. The simplest example of a single matrix element is that with the cross-spectral density function of a beam with separable phase: 
\begin{equation}
W^{(S)}(\pmb{\rho}_1,0,\pmb{\rho}_2,0;\omega)=R^{(S)}_l(\rho_1,\rho_2;\omega)e^{il(\phi_2-\phi_1)}.
\end{equation}
Then 
\begin{equation}
W^{(OAM)}_{ll}(\rho_1,\rho_2;\omega)=R^{(S)}_l(\rho_1,\rho_2;\omega),
\end{equation}
and all other elements are zero. The notable examples of this class are given in Refs. \cite{18} and \cite{36}. These sources are partially coherent, yet have a pure OAM state.

(B) \textit{Twisted Gaussian Schell-model (TGSM) sources}.  Twisted beams possess OAM that arises through a handedness built into the correlation properties of the beam, rather than through a coherent twist phase. The form of the TGSM CSD function is  \cite{19,37}
\begin{equation}\label{TwEx1}
\begin{split}
W^{(T)}(\pmb{\rho}_1,&0,\pmb{\rho}_2,0;\omega)=\exp\left[-\frac{\rho_1^2+\rho_2^2}{4\sigma^2}\right] \\&
\times \exp\left[-\frac{|\pmb{\rho}_1-\pmb{\rho}_2|^2}{2\delta^2}\right] \exp[i u(\pmb{\rho}_1 \times \pmb{\rho}_2)],
\end{split}
\end{equation}
where $u$ is a twist phase constrained to values $|u|\leq \delta^{-2}$. The CSD was shown to be represented as \cite{38,39}:
\begin{equation}\label{TwEx2}
\begin{split}
W^{(T)}(\pmb{\rho}_1,0,\pmb{\rho}_2,0;\omega)&=\sum\limits_{l=-\infty}^{\infty}R_l^{(T)}(\rho_1,\rho_2;\omega)e^{-il(\phi_1-\phi_2)}
\end{split}
\end{equation}
where
\begin{equation}\label{TwEx3}
\begin{split}
R_l^{(T)}(\rho_1,\rho_2;\omega)&=\exp\left[-\left(\frac{1}{4\sigma^2}+\frac{1}{2\delta^2}\right)(\rho_1^2+\rho_2^2)\right]  \\& \times\left(\sqrt\frac{1+u\delta^2}{1-u\delta^2} \right)^l I_l \left(\frac{\sqrt{1-u^2\delta^4}}{\delta^2} \rho_1 \rho_2 \right),
\end{split}
\end{equation}
$I_l$ being the modified Bessel function of the first kind and order $l$. Then 
\begin{equation}
W^{(OAM)}_{ll}(\rho_1,\rho_2;\omega)=R^{(T)}_l(\rho_1,\rho_2;\omega),
\end{equation}
while the off-diagonal elements vanish. In particular, if $u=0$, TGSM sources reduce to the classic Gaussian Schell-model sources. 

(C) \textit{Sources with radial coherence}. The general form of the source CSD with radial coherence was introduced in \cite{40}:
\begin{equation}
W^{(R)}(\pmb{\rho}_1,0,\pmb{\rho}_2,0;\omega)=f^*(\rho_1)f(\rho_2)g_{\phi}(\phi_1-\phi_2),
\end{equation}
where $g_{\phi}$ is a $2\pi$-periodic function of its argument. The quantity $W^{(R)}$ can be expressed as
\begin{equation}
W^{(R)}(\pmb{\rho}_1,0,\pmb{\rho}_2,0;\omega)=\sum\limits_{l=-\infty}^{\infty} g_l f^*(\rho_1)f(\rho_2)e^{-il(\phi_2-\phi_1)}, 
\end{equation}
and $g_n$ are the Fourier coefficients of function $g_{\phi}$: 
\begin{equation}
g_l=\frac{1}{2\pi}\int\limits_0^{2\pi} g_{\phi}(\phi)e^{-il\phi} d\phi.
\end{equation}
Hence the COAM matrix elements are
\begin{equation}
W^{(OAM)}_{ll}(\rho_1,0,\rho_2,0;\omega)=g_l f^*(\rho_1)f(\rho_2),
\end{equation}
and off-diagonal elements vanish.

(D) \textit{Incoherent superposition of Laguerre-Gaussian (LG) Schell-model beams}. Consider an incoherent sum of two LG modes, with topological charges -1 and 1 and Schell-like correlations.  Such a superposition can be constructed by creating two LG modes with different azimuthal orders, randomizing the beams separately, and then combining them. We then obtain $2\times2$ matrix 
\begin{equation}\label{IncEx1}
\begin{split}
W^{(I)}(\pmb{\rho}_1,0,\pmb{\rho}_2,0;\omega)&=\sum\limits_{l=\pm1}b_{l}U_l^*(\pmb{\rho}_1,0;\omega)U_l(\pmb{\rho}_2,0;\omega) \\& \times \mathrm{exp}\left[-\frac{\left|\pmb{\rho}_1-\pmb{\rho}_2\right|^2}{2\delta_{ll}^2}\right].
\end{split}
\end{equation}
where $\delta_{ll}$ is the correlation width between the fields with the same topological charge, and $b_l$ is the mode weight, 
\begin{equation}
U_{l}(\pmb{\rho},0;\omega)=C_l\rho^{\left|{l}\right|}\mathrm{exp}\left(il\phi\right)
\mathrm{exp}\left(-\frac{\rho^2}{w^2}\right),
\end{equation}
\begin{equation}
C_l=\sqrt{\frac{2}{\pi w^2\left|l\right|!}}\left(\frac{\sqrt{2}}{w}\right)^{\left|l \right|}, \quad  \{ l=-1,1 \}.
\end{equation}
where $w$ is the beam width. Decomposing the Gaussian correlation term in \eqref{IncEx1} into the infinite sum of helical Fourier modes, as done in Eqs. (\ref{TwEx2}) and (\ref{TwEx3}), yields
\begin{equation}
\begin{split}
W^{(I)}&(\pmb{\rho}_1,0,\pmb{\rho}_2,0;\omega)=C_1^2\rho_1\rho_2\exp\left[-\frac{\rho_1^2+\rho_2^2}{w^2}\right]\\&
\times \Biggl\{\sum_{l=-\infty}^{\infty} b_1 \exp\left[-\frac{\rho_1^2+\rho_2^2}{2\delta_{11}^2} \right]I_l\left(\frac{\rho_1\rho_2}{\delta_{11}^2}\right) \\& \times e^{i(l+1)(\phi_2-\phi_1)}\\&
+\sum_{l=-\infty}^{\infty} b_{-1} \exp\left[-\frac{\rho_1^2+\rho_2^2}{2\delta_{-1,-1}^2} \right]I_l\left(\frac{\rho_1\rho_2}{\delta_{-1,-1}^2}\right)\\& \times e^{i(l-1)(\phi_2-\phi_1)}
\Biggr\}.
\end{split}
\end{equation}  
On changing the summation indices as $l+1\rightarrow l$ and $l-1 \rightarrow l$ in the two sums, respectively, we can combine the terms with the same phase $\exp[il(\phi_2-\phi_1)]$. Hence the coefficients constitute the COAM matrix elements 
\begin{equation}
\begin{split}
W^{(OAM)}_{ll}&(\rho_1,0,\rho_2,0;\omega)=C_1^2\rho_1\rho_2\exp\left[-\frac{\rho_1^2+\rho_2^2}{w^2}\right]\\& 
\times \Biggl\{
b_1 \exp\left[-\frac{\rho_1^2+\rho_2^2}{2\delta_{11}^2} \right]I_{l-1}\left(\frac{\rho_1\rho_2}{\delta_{11}^2}\right)\\&
+ b_{-1} \exp\left[-\frac{\rho_1^2+\rho_2^2}{2\delta_{-1,-1}^2} \right]I_{l+1}\left(\frac{\rho_1\rho_2}{\delta_{-1,-1}^2}\right) \Biggr\}.
\end{split}
\end{equation}
This class of beams was recently shown to produce correlation-induced changes in the OAM flux density of the field \cite{41}, in analogy with correlation-induced spectral changes and correlation-induced polarization changes. The total OAM is conserved, but its density becomes redistributed throughout the cross-section of the beam on propagation.

All the sources considered above have COAM matrices with a diagonal structure, which indicates that most common partially coherent sources do not possess strong correlations between their OAM components. In fact, the only exception known to the authors relates to the surface plasmon polariton vortex fields  \cite{42}.

(E) \textit{Correlations via Bochner's theorem.} The next example introduces  a possible source with non-trivial off-diagonal elements. Let the $2\times2$ COAM matrix elements be
\begin{equation}\label{Ex}
\begin{split}
W_{p q}^{(OAM)}&(\rho_1,0,\rho_2,0;\omega)\\&= \frac{\Gamma(p/2+q/2)}{2c^{p/2+q/2}}\exp\left[- \frac{\rho_1^2+\rho_2^2}{4\sigma^2}\right] \rho_1^{p}\rho_2^{q}, 
\end{split}
\end{equation} 
where $p$ and $q$ can each take on the pair of values $\{p,q\}$. Here $\Gamma$ stands for Gamma function and $c>0$. This is a legitimate COAM matrix since it is evidently square-integrable with respect to $\rho_1$ and $\rho_2$ and quasi-Hermitian. To show that it is also non-negative definite it suffices to express $W(\pmb{\rho}_1,\pmb{\rho}_2;\omega)$ as the Bochner integral \cite{43}
\begin{equation}\label{Bochner}
W(\pmb{\rho}_1,\pmb{\rho}_2;\omega)=\int\limits_{-\infty}^{\infty}\int\limits_{-\infty}^{\infty} H^*(\pmb{\rho}_1,\textbf{v};\omega)p(\textbf{v};\omega)H(\pmb{\rho}_2,\textbf{v};\omega)d^2v, 
\end{equation}
where $\textbf{v}=(v,\psi)$ is a two-dimensional vector, $H(\pmb{\rho},\textbf{v};\omega)$ is a complex-valued function and $p(\textbf{v})\geq 0$. Indeed on setting
\begin{equation}
\begin{split}
H(\pmb{\rho},\textbf{v};\omega)&=\exp\left(- \frac{\rho^2}{2\sigma^2}\right) \\& \times \left[(\rho v)^{p}\exp(ip\phi)+(\rho v)^{q}\exp(iq\phi)\right]
\end{split}
\end{equation} 
and
\begin{equation}\label{EX2Ap}
p(v)=e^{-cv^2}, 
\end{equation} 
in \eqref{Bochner} and integrating results in $W_{p q}^{(OAM)}(\rho_1,\rho_2;\omega)$ as in \eqref{Ex}.

\begin{figure}[h]
    \centering
    \includegraphics[width=0.47\textwidth]{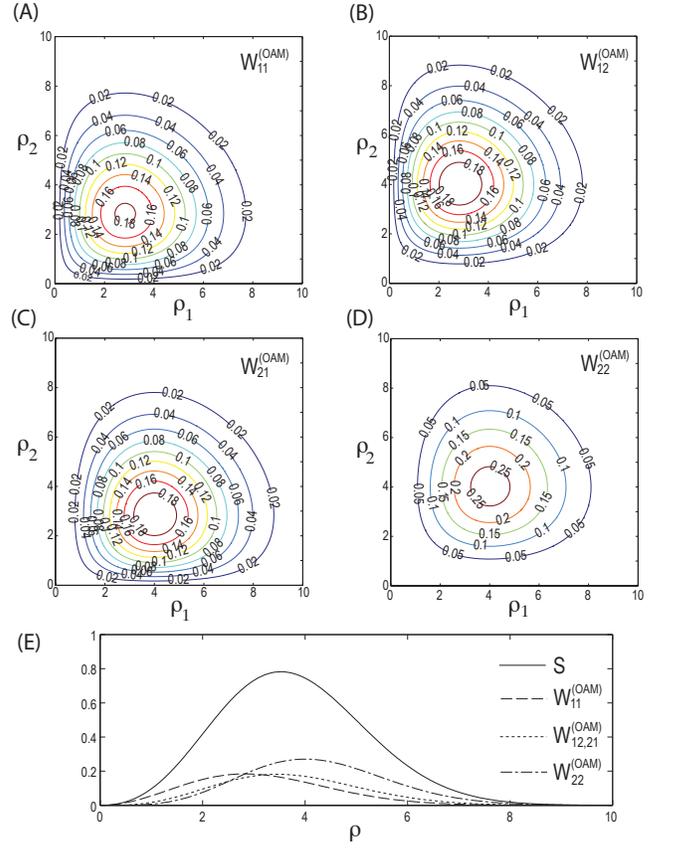}
    \caption{The COAM matrix elements (A) $W^{(OAM)}_{11}$; (B) $W^{(OAM)}_{12}$; (C) $W^{(OAM)}_{21}$; (D) $W^{(OAM)}_{22}$ varying with $\rho_1$ and $\rho_2$. (E) The COAM matrix elements at $\rho_1=\rho_2=\rho$ and $S(\rho)$ in \eqref{S3}. Parameters are $p=1$,  $q=2$, $\sigma=2$, $c=8$. $\rho_1$, $\rho_2$ and $\rho$ have any units of length, for example, mm. }
    \label{fig1}
\end{figure}

Figure \ref{fig1} presents the contour-plots of the COAM matrix elements in \eqref{Ex} with $p=1$ and $q=2$ depending on $\rho_1$ and $\rho_2$ in (A)-(D) as well as those at coinciding arguments $\rho_1=\rho_2=\rho$ and (E) the spectral density $S(\rho;\omega)$ given as
\begin{equation}\label{S3}
\begin{split}
S(\rho;\omega)&=W_{11}^{(OAM)}(\rho,\rho;\omega)+W_{12}^{(OAM)}(\rho,\rho;\omega)\\&+W_{21}^{(OAM)}(\rho,\rho;\omega)+W_{22}^{(OAM)}(\rho,\rho;\omega). 
\end{split}
\end{equation}

Figures \ref{fig1}(A) and \ref{fig1}(B) show that the diagonal elements of the COAM matrix are symmetric about line $\rho_1=\rho_2=\rho$, while the off-diagonal elements in \ref{fig1}(C) and \ref{fig1}(D) are not; however, they are reflections of each other about this line. This is the consequence of the quasi-Hermiticity  of the COAM matrix. Also, all four elements have different positions of maxima. Due to symmetry the off-diagonal elements coincide when $\rho_1=\rho_2=\rho$ [also shown in Fig. \ref{fig1}(E)]. By changing the values of parameters it appears possible to control the locations and values of maxima of the COAM matrix elements, however, those of the off-diagonal elements will be between those of the diagonal elements. It is possible to separate the four elements of the COAM matrix further apart but at the expense of the larger discrepancy between the maxima of the diagonal components. Such constrained behavior is the consequence of the non-negative definiteness.

(F) \textit{Partially correlated OAM states via unitary transformation.} We have noted that the quasi-Hermiticity and non-negative definiteness conditions listed in Section \ref{basicproperties} are independent of the specific choice of azimuthal basis. This observation implies that any unitary transformation of the COAM matrix will also produce a valid matrix. Considering that a matrix with diagonal elements can be constructed from the numerous examples of scalar cross-spectral densities, this strategy gives an easy method to construct non-trivial COAM matrices with correlations between OAM states. 

As a simple example, we begin with a COAM matrix with two pure states and no correlations between them. The matrix is of the form,
\begin{equation}
\begin{split}
\overleftrightarrow{W}&(\rho_1,0,\rho_2,0;\omega) \\&= \begin{bmatrix} W_1(\rho_1,0,\rho_2,0;\omega) & 0 \\ 0 & W_2(\rho_1,0,\rho_2,0;\omega) \end{bmatrix},
\end{split}
\end{equation}
where $W_1$ and $W_2$ are two distinct scalar cross-spectral densities. Now we employ a simple unitary transformation with the matrix,
\begin{equation}
\overleftrightarrow{M} \equiv \begin{bmatrix} \cos\psi & \sin\psi \\ -\sin \psi & \cos \psi \end{bmatrix},
\end{equation}
which results in a transformed matrix of the form $\overleftrightarrow{W}' = \overleftrightarrow{M} \overleftrightarrow{W} \overleftrightarrow{M}^\dagger$, or
\begin{equation}
\overleftrightarrow{W}'= \begin{bmatrix} W_1\cos^2\psi +W_2\sin^2\psi & (W_2-W_1)\sin2\psi/2 \\ (W_2-W_1) \sin2\psi/2  & W_1\sin^2\psi+W_2\cos^2\psi\end{bmatrix}.
\end{equation}
If we use this matrix in the OAM basis, it will represent a quasi-Hermitian, non-negative definite matrix with correlations between the two given OAM states. Matrices of any size with correlations between any OAM states can be introduced in a similar manner.

\section{Generalization to EM beams}

\label{EM:gen}

In this section we extend the COAM matrix to the electromagnetic domain and, hence, complete the coherence theory for all photonic states. We begin with Eq.~(\ref{SAM:elements}),  
\begin{equation}\label{SAM:again}
W_{\alpha \beta}(\Br_1,\Br_2;\omega) =\langle E^\ast_\alpha(\Br_1;\omega)E_\beta(\Br_2;\omega)\rangle_\omega,
\end{equation}
with the matrix represented in the circular polarization basis, and now decompose component $E_{\alpha}(\textbf{r};\omega)$ into the polar Fourier spectrum 
\begin{equation}\label{fielddecopmEM}
E_{\alpha}(\textbf{r};\omega)=E_{\alpha}(\rho,\phi,z;\omega)=\sum\limits_{l=-\infty}^{\infty} E_{\alpha l}(\rho,z;\omega)e^{i l\phi},
\end{equation}
where Fourier coefficients, being functions of $\rho$ and $z$ are
\begin{equation}\label{FTcoefficientsEM}
E_{\alpha l}(\rho,z;\omega)=\frac{1}{2\pi}\int\limits_0^{2\pi}E_{\alpha}(\rho,\phi,z;\omega)e^{-i l\phi}d\phi.
\end{equation} 
On substituting from \eqref{fielddecopmEM} into \eqref{SAM:again} and writing the result as a matrix that we term \textit{Coherence OAM SAM [COS]} matrix 
\begin{equation}\label{COS1}
\begin{split}
\overleftrightarrow{W}^{(COS)}&(\rho_1,z_1,\rho_2,z_2;\omega)\\&=[W_{l\alpha  \beta m}^{(COS)}(\rho_1,z_1,\rho_2,z_2;\omega)] \\&
=[\langle E^*_{\alpha l}(\rho_1,z_1;\omega)  E_{\beta m}(r_2,z_2;\omega) \rangle], \\&
 \quad  (\alpha,\beta =+,-), \quad (-\infty < l,m < \infty).
\end{split}
\end{equation}
whose individual elements have form
\begin{equation}\label{COS2}
\begin{split}
&W_{l\alpha  \beta m}^{(COS)}(\rho_1,z_1,\rho_2,z_2;\omega)\\&=\frac{1}{(2\pi)^2} \int\limits_0^{2\pi}\int\limits_0^{2\pi} e^{i l\phi_1} W_{\alpha\beta}(\textbf{r}_1,\textbf{r}_2;\omega)  e^{-i m\phi_2}d\phi_1 d\phi_2, \\&
\quad   \quad\quad\quad\quad\quad\quad\quad\quad\quad   -\infty <l,m< \infty.
\end{split}
\end{equation}
We have written this matrix as depending on four indices -- $l \alpha  \beta m$ -- but it can readily be reorganized as a single matrix. One can consider it an OAM matrix with an SAM sub-matrix for each element, or an SAM matrix with an OAM sub-matrix for each element. Further work will be needed to determine which description is most convenient. 

Many of the properties derived for the scalar COAM matrix can be generalized for the full electromagnetic case. For example, using the definitions of the orbital angular momentum flux density and the spin flux density for partially coherent fields from Ref.~\cite{14}, we may write the orbital flux density as
\begin{eqnarray}
M_{o}(\textbf{r};\omega) &=& \frac{\varepsilon_0}{2k}Im[\partial_{\phi_2}W_{++}(\rho_1,\phi_1,z,\rho_2,\phi_2,z;\omega)  \nonumber \\
&+&\partial_{\phi_2}W_{--}(\rho_1,\phi_1,z,\rho_2,\phi_2,z;\omega)]_{r_1=r_2}, \nonumber \\
\end{eqnarray}
and the spin flux density as
\begin{eqnarray}
M_{s}(\textbf{r};\omega) &=& \frac{\varepsilon_0}{2k}Re[ W_{++}(\rho_1,\phi_1,z,\rho_2,\phi_2,z;\omega) \nonumber \\ 
&-& W_{--}(\rho_1,\phi_1,z,\rho_2,\phi_2,z;\omega)]_{r_1=r_2},
\end{eqnarray}
On applying our definition of the COS matrix, we have
\begin{equation}\label{M:orbit}
\begin{split}
M_{o}(\textbf{r};\omega)& = \frac{\varepsilon_0}{2k} \sum_{\alpha=+,-}\Biggl[ \sum\limits_{l=-\infty}^{\infty} lW_{l\alpha\alpha l}^{(COS)}(\rho,z,\rho,z;\omega)\\& 
+  \sum\limits_{l,m=-\infty}^{\infty}Re [m W_{l \alpha\alpha m}^{(COS)} (\rho,z;\omega) e^{-i(l-m)\phi}]
\Biggr], 
\end{split}
\end{equation}
while the spin flux density has the form,
\begin{equation}\label{M:spin}
\begin{split}
M_{s}(\textbf{r};\omega)& = \frac{\varepsilon_0}{2k}\sum_{\alpha=+,-} \alpha \Biggl[ \sum\limits_{l=-\infty}^{\infty} W_{l\alpha\alpha l}^{(COS)}(\rho,z,\rho,z;\omega)\\& 
+  \sum\limits_{l,m=-\infty}^{\infty}Re [ W_{l \alpha\alpha m}^{(COS)} (\rho,z;\omega) e^{-i(l-m)\phi}]
\Biggr], 
\end{split}
\end{equation}
Other local and nonlocal coherence and OAM properties can be derived in a similar manner.

\section{Summary}

We have developed a theoretical treatment for characterization of the second-order correlation properties of stationary optical beam-like fields in all possible OAM and SAM states. The central quantity of our theory is the Coherence-OAM (COAM) matrix whose elements describe scalar correlations between the fields having any two azimuthal orders, say $l$ and $m$. This matrix is obtained by first filtering the complex Fourier series components of the field realizations with respect to the angular variable and then correlating the obtained Fourier coefficients of the series, which are functions of radial variable. The COAM matrix is shown to be (generally) infinitely-dimensional, quasi-Hermitian and non-negative definite, among other properties. Important particular cases of the OAM matrix have been discussed: pure state (single diagonal element), fully coherent state (factorizable with the same coefficient) and fully correlated state (factorizable with different coefficients).  

We have established that the local values of the spectral density, the OAM density flux and the degree of coherence of the scalar beam depend on both diagonal and off-diagonal elements of the COAM matrix but the cross-section integrated values only depend on  diagonal elements. A new quantity characterizing the degree of OAM purity is also introduced and two definitions for it were suggested, one in similarity with the degree of polarization and the other in similarity with the von Neumann's entropy.

We have also obtained the free-space Wolf equations for the COAM matrix elements in paraxial regime in form of double integrals over the source radial variables, using the Hankel transform method. The form of such propagation laws depends on values of $l$ and $m$. Since the COAM matrix elements propagate independently but each depend on their (generally different) distributions in the source plane, the Wolf equations can be used to characterize the correlation-induced OAM changes \cite{41}. This new effect is in line with the previously known source correlation-induced spectral and polarization changes.
 
We have proposed a modified version of the Young's double slit interference experiment for resolving the individual components of the COAM matrix. The experimental setup involves a pair of concentric rings rather than slits arranged in the opaque screen and uses linear phase spiral plates behind them. The COAM matrix element with indices $l$ and $m$ is then obtained in the focal region of the lens placed just behind the opaque screen, as the charges of the spiral plates are fixed at $l$ and $m$ and the radii of the rings let vary from zero to values sufficiently larger than the beam extent.   

In order to show the usefulness of the theory we have overviewed several existing families of stationary beams carrying OAM, including those with a separable phase and with a twist, and calculated their COAM matrices. The calculation revealed that all the existing models are described with diagonal OAM matrices. As the last two examples we have introduced stationary beams with genuine COAM matrices having non-trivial off-diagonal elements, the possibility not previously known.  

Finally, we have merged the COAM matrix treatment for scalar fields with the previously known Coherence-SAM matrix (cross-spectral density matrix in circular polarization basis) treatment for electromagnetic beams. The result is the Coherence OAM-SAM matrix capable of characterizing the second-order correlations in stationary fields having any polarization and OAM states. Our theory was developed in space-frequency domain and is valid for beams with any spectral composition. It is also possible to develop a parallel space-time formulation, on resorting to  quasi-monochromatic fields.    

\section*{APPENDIX A: Derivation of the Wolf equations for the COAM matrix}

\label{wolfeqn}

A scalar optical field $U(\textbf{r};\omega)$ propagating along the $z$-direction satisfies the Helmholtz equation
\begin{equation}\label{HelmholzE}
\nabla^2 U (\textbf{r};\omega)+k^2 U(\textbf{r};\omega)=0, 
\end{equation}
where $\nabla^2$ denotes the Laplacian operator and $k=\omega/c$, $c$ being the speed of light. In cylindrical coordinates $\textbf{r}=(\rho,\phi,z)$, the Laplacian takes the form
\begin{equation}\label{Lap}
\begin{split}
\nabla^2U&(\rho,\phi,z;\omega)\\& =\Bigl[\frac{\partial^2}{\partial \rho^2}+\frac{1}{\rho}\frac{\partial}{\partial \rho} +\frac{1}{\rho^2}\frac{\partial^2}{\partial \phi^2}+\frac{\partial^2}{\partial z^2}\Bigr]U(\rho,\phi,z;\omega).
\end{split}
\end{equation}
On applying the Fourier series representation (\ref{fielddecopm1}) in \eqref{HelmholzE} while using the Laplacian in \eqref{Lap}, we arrive at equation
\begin{equation}
\sum\limits_{l=-\infty}^{\infty}  \Bigl[
\frac{\partial^2}{\partial \rho^2}+\frac{1}{\rho}\frac{\partial}{\partial \rho} +\left(k^2-\frac{l^2}{\rho^2}\right)+\frac{\partial^2}{\partial z^2}
\Bigr]U_l(\rho,z;\omega)e^{il\phi}=0
\end{equation} 
In order for this equation to hold, it must be satisfied term by term for all values of $l$, i.e.,
\begin{equation}\label{fieldEq}
\begin{split}
\Bigl[\frac{\partial^2}{\partial \rho^2}+\frac{1}{\rho}\frac{\partial}{\partial \rho} +\left(k^2-\frac{l^2}{\rho^2}\right)+\frac{\partial^2}{\partial z^2}\Bigr]&U_l(\rho,z;\omega)=0, \\&
  (-\infty< l<\infty).
\end{split}
\end{equation}

We are most interested in the propagation of beam-like fields in the paraxial approximation. We derive a paraxial form of the propagation equation by substituting
\begin{equation}
U_l(\rho,z;\omega)=u_l(\rho,z;\omega)e^{ikz}
\end{equation}
into \eqref{fieldEq}. Then on canceling $e^{ikz}$ throughout and neglecting term $\frac{\partial^2}{\partial z^2} u_l(\rho,z;\omega)$ we obtain equations
\begin{equation}\label{parax}
\begin{split}
\left[\frac{\partial^2}{\partial \rho^2}+\frac{1}{\rho}\frac{\partial}{\partial \rho} -\frac{l^2}{\rho^2}+2ik\frac{\partial}{\partial z}\right]&u_l(\rho,z;\omega)=0, \\&
  (-\infty< l<\infty).
\end{split}
\end{equation}
By considering correlations of different field components, 
\begin{equation}\label{W:parax}
w_{lm}^{(OAM)}(\rho_1,z_1,\rho_2,z_2;\omega)=\langle u_l^*(\rho_1,z_1;\omega)u_m(\rho_2,z_2;\omega) \rangle, 
\end{equation}
we arrive at the pair of equations
\begin{equation}\label{WE2}
\begin{split}
\Bigl[\frac{\partial^2}{\partial \rho_i^2}+\frac{1}{\rho_i}\frac{\partial}{\partial \rho_i} & -\frac{l^2}{\rho_i^2}+2ik\frac{\partial}{\partial z_i}\Bigr]\\& \times w^{(OAM)}_{lm}(\rho_1,z_1,\rho_2,z_2;\omega)=0,
\\&  (-\infty <l,m<\infty) , \quad(i=1,2). 
\end{split}
\end{equation}
Equations (\ref{WE2}) represent the Wolf equations for the COAM matrix elements in the paraxial approximation, describing the evolution of the correlation functions. In order to obtain the solution of these equations, we first find the solution for paraxial field $u_l(\rho,z;\omega)$ by the use of the integral transform method. Namely, we first take the Hankel transform of order $l$, defined as \cite{44}
\begin{equation}
\mathcal{H}_l[u_l(\rho,z;\omega)]=\hat{u}_l(\kappa,z;\omega)=\int\limits_0^{\infty}\rho J_l(\kappa \rho)u_l(\rho,z;\omega)d\rho,
\end{equation}
with $J_l$ being the Bessel function of the first kind and order $l$, of \eqref{parax}. Then, by means of the identity
\begin{equation}
\mathcal{H}_l\left[\left(\frac{\partial^2}{\partial \rho^2}+\frac{1}{\rho}\frac{\partial}{\partial \rho} -\frac{l^2}{\rho^2}\right) u_l(\rho,z;\omega) \right]=-\kappa^2 \hat{u}_l(\kappa,z;\omega),
\end{equation}  
it takes the form
\begin{equation}
\kappa^2 \hat{u}_l(\kappa,z;\omega)+2ik\frac{\partial}{\partial z}\hat{u}_l(\kappa,z;\omega)=0. 
\end{equation}
This first-order ordinary differential equation can be readily solved to yield
\begin{equation}\label{Hsolution}
\hat{u}_l(\kappa,z;\omega)=\exp\left(-i\frac{\kappa^2 z}{2k} \right)\hat{u}_l(\kappa,0;\omega),
\end{equation}
where 
\begin{equation}\label{Hinitial}
\hat{u}_l(\kappa,0;\omega)=\int\limits_0^{\infty}\rho' J_l(\kappa \rho')u_l(\rho',0;\omega)d\rho'
\end{equation} 
is the Hankel transform of the field in the source plane. On applying the inverse Hankel transform of \eqref{Hsolution}: 
\begin{equation}
\mathcal{H}^{-1}_l[\hat{u}_l(\kappa,z;\omega)]=u_l(\rho,z;\omega)=\int\limits_0^{\infty}\rho J_l(\kappa \rho)\hat{u}_l(\kappa,z;\omega)d\kappa,
\end{equation} 
then using \eqref{Hinitial} and changing the order of integration, we arrive at the solution
\begin{equation}\label{paraxprop}
\begin{split}
u_l(\rho,z;\omega)&=\int\limits_0^{\infty}\rho' u_l(\rho',0;\omega)d\rho'\\&
\times \left[\int\limits_0^{\infty}\kappa J_l(\kappa \rho) J_l(\kappa \rho') \exp\left(-i\frac{\kappa^2 z}{2k} \right) d\kappa \right].
\end{split}
\end{equation}

On correlating the fields on the left and the right sides of  \eqref{paraxprop}, we obtain the Huygens-Fresnel integral for the COAM matrix elements:
\begin{equation}
\begin{split}
&w^{(OAM)}_{lm}(\rho_1,z_1,\rho_2,z_2;\omega)\\&=\int\limits_{0}^{\infty}\int\limits_{0}^{\infty} \Biggl[ \rho'_1 \rho'_2 w^{(OAM)}_{lm}(\rho_1,0,\rho_2,0;\omega) \\& \times
\int\limits_0^{\infty}\kappa_1 J_l(\kappa_1 \rho_1) J_l(\kappa_1 \rho'_1) \exp\left(i\frac{\kappa_1^2 z_1}{2k} \right) d\kappa_1 \\& \times
\int\limits_0^{\infty}\kappa_2 J_l(\kappa_2 \rho_2) J_l(\kappa_2 \rho'_2) \exp\left(-i\frac{\kappa_2^2 z_2}{2k} \right) d\kappa_2\Biggr] d\rho'_1d\rho'_2. 
\end{split}
\end{equation}

\subsection*{Acknowledgements} Olga Korotkova acknowledges the support from the University of Miami via the Cooper Fellowship program.

\subsection*{Funding} Office of Naval Research of America (No.~MURI N00014-20-1-2558).

\qquad

\end{document}